\documentclass{article}
\usepackage{amsmath}
\usepackage{graphicx}
\usepackage{subcaption}

\begin{document}

\centerline{\LARGE \bf Forecasting the distribution of long-horizon returns}

\vskip2mm

\centerline{\LARGE \bf  with time-varying volatility}

\vskip6mm

\centerline{\large  Hwai-Chung Ho}

\vskip2mm
\centerline {\large  Academia Sinica and National Taiwan University }

\vskip2mm



\date{}

\vskip1cm


\section{Background}

The study of long-horizon returns has received a great deal of attention in recent years (see, for example, Boudoukh, Richardson, and Whitelaw (2008), Neuberger (2012) and Lee (2013), Fama and French (2018)). While most of the discussions are concerned with some practical issues in investment, few have touched the important aspect on risk
management. The approach adopted in this article is to  predict the future distribution of  the returns of a fixed long-horizon by which the risk measures of interest that come in the form of a distributional functional such as the value at risk (VaR) and the conditional tail expectation (CTE) can be easily derived.
The characteristic feature of our approach which requires no specification of the volatility dynamics nor parametric assumptions of the shock distribution extends the work by Ho et al. (2016) and Ho ( 2017) to a more general volatility dynamics that includes both the widely-used SV model and  the GARCH model (Bollerslev, 1986) as special cases. \\



\section{Model} \label{model}

We consider a general time-varying  volatility (GV) model for the return $r_{t}$  which is in the form of
\begin{equation}
r_{t}=\mu + h_{t}u_{t}, \quad h_t = f(\eta_{t-1}, \eta_{t-2}, \ldots, ) \label{model},
\end{equation}
and satisfies the following two conditions.

 (i)  $\{u_t\}$ is the sequence of iid symmetric shocks with zero mean and unit variance.

(ii) $\{\eta_i, -\infty < i < +\infty\}$ is a strictly stationary sequence independent of $\{u_t\}$, and $f$ is a positive measurable function  such that $\sigma^2=$ $Eh_t^2 =$ $E(r_t - \mu)^2 < \infty$.   \\

The GV model defined in  ({\ref{model}) generalizes a host of popular models proposed in the literature for financial econometrics, in particular, the stochastic volatility model (SV) and the ARCH-type model.\\

\noindent (i) If $\{\eta_t\}$ is an iid sequence independent of the normal shocks $\{u_t\}$, and $f$ is
\begin{equation}
f(x_1, x_2,\ldots,)= \delta_0 \exp \left\{\sum_{i=0}^{\infty} \phi^i x_i /2 \right\}, \quad 0 < \phi < 1, \nonumber \label{sv}
\end{equation}
then  (\ref{model}) represents the standard SV model (Taylor, 1994).\\

\noindent (ii) Suppose the volatility component $h_t$ follows the square GARCH(p,q) dynamic equation (Bollerslev, 1986)
\begin{equation*}
h_t^2 = \alpha_0 + \sum_{i=1}^p \alpha_i h_{t-i}^2 + \sum_{j=1}^q \beta_j \left( r_{t-j} - \mu \right)^2. \label{garch}
\end{equation*}
By using the shift-operator method, it is not difficult to see that $h_t^2$ can be expressed as an infinite moving
average of $(r_t-\mu)^2$'s, i.e.,
\begin{equation}
h_t^2 = \sum_{i=1}^{\infty} g_i \left( r_{t-i} - \mu \right)^2 \nonumber
\end{equation}
for some summable sequence  $\{g_i\}$. Equation (\ref{model}) hence covers the case of GARCH (p,q) sequence with
$f(x_1, x_2,\ldots) = ( \sum_{i=0}^{\infty} g_i x_i  )^{1/2}$ and $\eta_{t-1} = (r_{t-1} -\mu)^2$.\\


\section{Forecasting distribution} \label{fd}

 Let $\{r_t\}$ be the underlying return sequence modeled by (\ref{model}). For a
fixed $T$, let $R_{t,T}=$ $\sum_{s=t+1}^{t+T } r_s$ denote the integrated returns of horizon $T$ from $t+1$ to $t+T$,  and let $F^{R}_{T}(\cdot | {\cal F})$ be the distribution of $R_{N,T}$ conditional on the information set ${\cal F}$ generated by a set of past returns. We replace $F^{R}_{T}(\cdot | {\cal F})$  by $F^{R}_{N,T}(\cdot )$ when ${\cal F}$ is generated by $\{r_s,  1 \le s \le N\}$

Model (\ref{model}) entails an interesting observation. Let $\Psi_t$ denote the information set generated by $\{|r_i-\mu|, 1 \le i \le t\} \}$, and define the sign function
$\pi_i$ of the shock $u_i$, i.e., $\pi_i = u_i / |u_i|$. Let $(a_1, \ldots, a_t)$ $\in$ $\{-1, 1\}^t$ with $a_i$ $= -1$ or $1$ be an outcome of the random vector
$(\pi_1, \ldots, \pi_t)$. Set $I_i = (0, +\infty)$ if $a_i =1$, or $(-\infty, 0)$ if $a_i = -1$. Then
\begin{eqnarray}
&& P\left( (\pi_1, \ldots, \pi_t) = (a_1, \ldots, a_t) | \Psi_t \right) \nonumber \\
&& = P\left( (\pi_1, \ldots, \pi_t) = (a_1, \ldots, a_t) | |u_1|, \ldots, |u_t| \right)  \nonumber \\
&& = P\left(u_1 \in I_1, \ldots, u_t \in I_t | |u_1|, \ldots, |u_t|  \right), \nonumber \\
&& = P\left(u_i \in (0, +\infty), 1 \le i \le t | |u_1|, \ldots, |u_t|  \right), \label{property} \\
&& = \Pi_{i=1}^t P\left(u_i \in (0, +\infty) \right), \nonumber \\
&& = 2^{-t}  \nonumber \\
&& = \Pi_{i=1}^t P\left(\pi_i = a_i \right), \nonumber
\end{eqnarray}
where the third equality holds due to the assumption of symmetric $u_t$. We formalize (\ref{property}) in the following statement.\\

\noindent {\bf Property I}. Conditional on $\Psi_t$, $\left\{\pi_i, 1 \le i \le t \right \}$  forms an iid Rademacher sequence with $P( \pi_i = 1) = P (\pi_i = -1) = 1/2$. \\

Suppose for the time being $\mu$  of $r_t$ in (\ref{model}) is known.  Let  $r_t - \mu = m_t \delta_t$, where $m_t = |r_t - \mu|$ and $\delta_t  = 1$ or $-1$ is the sign of  $r_t - \mu$.
Given $\{m_t,  1 \le t \le N \}$,  denote by  $\widehat{m}_{n, N, h}$ with $h \ge 1$  the best linear predictor of $m_{N+h}$ based on  the past $n$ returns of $\{m_t,  N-n+1 \le t \le N \}$, and let $\Delta_h (N, n)  = m_{N+h} - \widehat{m}_{n, N, h}$ be the prediction error. Write
\begin{eqnarray}
R_{N, T} - T \mu & = & \sum_{h=1}^{T} \left(r_{N+h} - \mu \right) \nonumber\\
                           & = & \sum_{h=1}^{T} m_{N+h} \delta_{N+h} \label{delta}  \\
                          &= &   \sum_{h=1}^{T}  \left( \widehat{m}_{n,N,h} + \Delta_h (N,n) \right) \delta_{N+h}. \nonumber
\end{eqnarray}
From {\bf Property I} we know that $\{\delta_t\}$ is an iid Rademacher sequence.  What left to be specified in order to approximate  $m_{N+h} \delta_{N+h}$ is the $\Delta_h (N,n)$ on the right-hand side of the last equation in (\ref{delta}).  Because  $\Delta_h (N,n)$ is the prediction errors  of future observations (with $N$ being the current time) that are perpendicular to $\{ m_t, 1 \le t \le N\}$ , it would not be feasible to predict their values. Yet the distribution of $\Delta_h (N,n)$ for different $h$'s can be retrieved empirically from the past prediction errors. We therefore propose using  $ \widehat{m}_{n,N,h} + W_{n,h}$ as the proxy for the forecast of $m_{N+h}$ where $W_{n,h}$ is an independent copy of the $h$-step prediction error $\Delta_h (t,n) = m_{t+h} - \hat{m}_{n,t,h} $ for fixed $n$ and $h$. We choose $W_{n,h}$ not only because it is identically distributed with $\Delta_h (N,n)$ but also that the two random vectors $(\widehat{m}_{n,N,1} + W_{n,1}, \ldots, \widehat{m}_{n,N,T} + W_{n,T})$ and  $(\widehat{m}_{n,N,1} + \Delta_1 (N,n), \ldots, \widehat{m}_{n,N,T} + \Delta_T (N,n))$ have the same covariance matrix, since
\begin{eqnarray}
  &&\mbox{cov}\left(\widehat{m}_{n,N,h} + W_{n,h}, \widehat{m}_{n,N,k} + W_{n,k} \right)\nonumber \\
   & & = \mbox{cov}\left(\widehat{m}_{n,N,h} + \Delta_h (N,n), \widehat{m}_{n,N,k} + \Delta_k (N,n)\right) \nonumber \\
   & & = \mbox{cov} \left(\widehat{m}_{n,N,h}, \widehat{m}_{n,N,k} \right) \nonumber
 \end{eqnarray}
for $h, k = 1, \ldots, T$. To construct the best $h$-step linear predictor $\hat{m}_{n,N,h}$ for $m_{N+h}$ based on $\{r_t, N-n+1 \le t \le N\}$, we adopt the widely used innovation algorithm which only requires  the covariance function of $\{m_t\}$. More specifically, let $m_t^* = m_t - \mu_m$ where $E m_t = \mu_m$, then the one-step predictor  $\widehat{m}_{n, t, 1}^*$ of  $m^*_{t+1}$  based on $\{m^*_s, t-n+1 \le s \le t\}$ can be expressed as
\begin{equation}
\widehat{m}_{n, t, 1}^* = \sum_{j=1}^n \theta_{n,j} \left(m^*_{t+1-j}  - \widehat{m}^*_{n-j, t-j,1} \right), \label{one-step}
\end{equation}
where the coefficients $\theta_{n,j}$ are determined by a set of recursive equations built on the covariances of $\{m^*_t\}$ (Proposition 5,2,2, Brockwell and Davis, 1991). In (\ref{one-step})
we set $\widehat{m}^*_{ x, y, z} $  to be zero if  $x=0$.  For $h$-step prediction, it follows from equation (\ref{one-step}) that
\begin{equation}
\widehat{m}_{n, t, h}^* = \sum_{j=h}^{n-h+1} \theta_{n+h-1,j} \left(m^*_{t+1-j}  - \widehat{m}^*_{n-j, t-j,1} \right), \label{h-step}
\end{equation}
using the fact that the prediction error $m^*_{t+h-j}- \widehat{m}^*_{n-j, t+h-j,1}$  for $h -j \ge 1$ is othogonal to the linear span of $\{m^*_s, t-n+1 \le s \le t\}$  (Section 5.2, Brockwell and Davis, 1991).  Note that  in (\ref{one-step}) and (\ref{h-step}) the underlying time series $\{m_t^*\}$ for which the predicton is carried out is of mean zero. Thus, it is necessary to add  the mean back to $\widehat{m}_{n, t, h}^* $ to obtain  the predictor $\widehat{m}_{n,t,h} =  \widehat{m}_{n, t, h}^* + \mu_m$ for the observation ${m}_{t+h}$. When the mean $\mu_m$ is unknown, we modify the predictor to be $\widehat{m}_{n,t,h} =  \max\{\widehat{m}_{n, t, h}^* + \widehat{\mu}_m, 0 \}$ to conform to the non-negativity of $m_t$.  For $h = 1, \ldots, T$, let ${\cal E}_h $ be the set of all the observed prediction errors  $\Delta_h (t,n)$ created by the past observations, that is,
\begin{equation}
{\cal E}_h  (N,n) = \{ \Delta_h (t,n) = m_{t+h} - \hat{m}_{n,t,h}, t = n,  n+1, \ldots, N\}. \label{eh}
\end{equation}
Define
\begin{equation}
\widehat{R}_{n,N, T} = T\hat{\mu} +  \sum_{h=1}^{T}  \left( \widehat{m}_{n,N,h} + \widehat{W}_{n,h} \right)\delta^*_{h} \label{rnt}
\end{equation}
where $\widehat{W}_{n,h}$ is the random variable designed to approximate $W_{n,h}$ and  having the same distribution as the empirical distribution constructed by the elements of  ${\cal E}_h (N,n) $, and  $\{\delta^*_{h}, h = 1,\ldots, T\}$ is an iid Rademacher sequence independent from $\{m_t\}$.  For $h=1, \ldots, T$
we draw independen samples  $\{\widehat{W}_{n,h}^{(i)}, h=1, \ldots, T \}$  and  $ \{\delta^{*(i)}_{h}, h=1, \dots, T\}$ with the former from   ${\cal E}_h (N,n)$  to form
\begin{equation}
\widehat{R}^{(i)}_{n, N, T} = T\hat{\mu} +  \sum_{h=1}^{T} \left( \widehat{m}_{n,N,h} + \widehat{W}_{n,h}^{(i)} \right) \delta^{*(i)}_{h}. \label{rnti}
\end{equation}
Then the empirical distribution  $\widehat{F}^{R}_{N,T}(\cdot )$    built on  $ \{\widehat{R}^{(i)}_{N, T}, i = 1,\ldots,B \} $ for a large $B$ is the forecast we propose for the distribution   $F^{R}_{N,T}(\cdot )$  of $R_{N,T}$  conditional on $\{r_t,  1 \le t \le N\}$. \\


\section{Presence of stochastic trends} \label{pt}

The GV model previously described gives a white noise sequence which is a  common stylized facts exhibited by the returns of market indexes (Asset Price Dynamics, Volatility, and Prediction by SJ Taylor, 2005).  For individual stocks, however, the return sequence  tends to show serial correlations. In this section we extend the forecasting method discussed in Section \ref{fd} to the case where the returns are serially correlated. For ease of presentation, we from now on use $x_t$ to denote the return and $S_{N, T} = \sum_{t=N+1}^{N+T} x_t$ for its integrated returns. Let   $\widehat{F}^{S}_{N,T}(\cdot )$ be the distribution of $S_{N,T}$ conditional on $\{x_t, 1 \le t \le N \}$.  Assume that $x_t$ is invertible with respect to  $\{ r_t\}$, that is, for some smooth function $f (x) = \sum_{i=0}^{\infty} a_i x^i$ with $a_0 =1$,
\begin{equation}
r_t = f(B)(x_t - \mu) = \sum_{i=0}^{\infty} a_i \left(x_{t-i} - \mu \right),  \label{rt}
\end{equation}
 where $B$ is the shift operator and $\{r_s\}$ is a zero-mean innovation sequence following  the GV model. Model (\ref{rt}) includes the familiar ARMA process with ARCH-type or SV innovations. Given a sample $\{ x_t, 1 \le t \le N\}$ of size $N$,  similar to the notation used in Section \ref{fd},
set $x^*_t = x_t -\mu$ and let $\hat{x}^*_{n, N, h}$    denote the best linear $h$-step predictor of $x^*_{N+h}$ base on $\{ x^*_t, N-n+1 \le t \le N\}$.  Using the innovation algorithm, we express each future $x^*_{N+h}$ as a weighted sum of prediction errors,
\begin{eqnarray}
x^*_{N+h} & = &  \left(x^*_{N+h}  - \hat{x}^*_{n, N+ h-1,1} \right) +  \hat{x}^*_{n, N+h-1, 1} \nonumber \\
                   &=&   \sum_{j=0}^n \theta_{N+h-1, j} \left( x^*_{N+h-j} - \hat{x}^*_{n, N+h-1-j, 1} \right) \nonumber \\
                 & = &  \sum_{j=h}^n \theta_{N+h-1, j} \left(x^*_{N+h-j} - \hat{x}^*_{n, N+h-1-j, 1} \right) \\
                  & & \quad + \sum_{j=0}^{h-1}  \theta_{N+h-1, j}  \left(x^*_{N+h-j} - \hat{x}^*_{n, N+h-1-j, 1} \right),  \nonumber
\end{eqnarray}
where $\theta_{\cdot, 0} =1$. Then we write the integrated return $S_{N, T}$   of $x^*_{t}$ as
\begin{eqnarray}
S_{N, T} -T\mu  &=& \sum_{h=1}^{T} x^*_{N+h} \nonumber \\
                 & = &  \sum_{h=1}^T \sum_{j=h}^n \theta_{N+h-1, j} \left(x^*_{N+h-j} - \hat{x}^*_{n, N+h-1-j, 1} \right ) \nonumber  \\
                  & & \quad + \sum_{h=1}^T \sum_{j=0}^{h-1}  \theta_{N+h-1, j} \left(x^*_{N+h-j} - \hat{x}^*_{n, N+h-1-j, 1} \right) \nonumber \\
                 & \equiv & {\cal T}_{n, N, T} + {\cal I}_{n, N,T} \label{snt}
\end{eqnarray}
 By grouping the coefficients $\theta_{i,j}$ that associated with the same estimated innovation term $x^*_{N+h-j}- \hat{x}^*_{n, N+h-j-1,1}$, we can write $ {\cal T}_{n, N, T}$ and  ${\cal I}_{n, N,T}$ as
\begin{equation}
 {\cal T}_{n, N, T}   =   \sum_{j=1}^n \left( \sum_{h=1}^T  \theta_{N+h-1, j} \right)  \left( x^*_{N+1-j} - \hat{x}^*_{n, N-j, 1} \right), \nonumber
 \end{equation}
\begin{equation}
{\cal I}_{n, N,T}   =   \sum_{j=N+1}^{N+T} \left( \sum_{h=0}^{n+T-j}  \theta_{j+h-1, h} \right) \left( x^*_{j} - \hat{x}^*_{n, j -1, 1} \right). \nonumber
 \end{equation}
In the decomposition (\ref{snt}) of $S_{N,T}$, ${\cal T}_{n, N,T}$  represents the linear forecast of  $S_{N,T} - T\mu$  derived from the given sample $\{x^*_t, N-n+1 \le t \le N\}$,  and  ${\cal I}_{n, N,T}$ consists of  the one-step prediction errors  not directly observed.  Thus an acceptable candidate for forecasting the distribution  of $S_{N,T}$  would  be in the form of  ${\cal T}_{n, N,T} +  \widehat{F}^{I}_{N,T}(x )$ where  $\widehat{F}^{I}_{N,T}(x )$ is a good estimate of the  conditional distribution  $F^{I}_{N,T}(x )$  of  ${\cal I}_{n, N,T}$ given  $\{x^*_t, 1 \le t \le N\}$.  Before we proceed to find   $\widehat{F}^{I}_{N,T}(x )$,  we first note that due to assumption (\ref{rt})  the one-step prediction error $x^*_{t} - \hat{x}^*_{n, t-1, 1} $ is close to $ r_{t}$ for each $t$ if $n$ is sufficiently large. Therefore we may regard  ${\cal I}_{N,h}$ as a weighted sum of uncorrelated $r_t$'s. Define $z_j(N,n) = x^*_{j} - \hat{x}^*_{n, j -1, 1}$, $\widetilde{m}_j (N,n) = |z_j (N,n)|$, and $\widetilde{\delta}_j  (N,n) = z_j(N,n)/ \widetilde{m}_j (N,n)$.
For $j = N+1, \dots, N+T$, let $\widehat{\widetilde{m}}_j (N,n)$ denote the best linear forecast of $\widetilde{m}_j (N,n)$ based on the previous prediction errors $\{ \widetilde{m}_t (N,n) = |x^*_{t} - \hat{x}^*_{n, t -1, 1}|, n+1 \le t \le N \}$.

Similar to finding an estimate for the conditional distribution $F^{R}_{N,T} (\cdot)$ as presented in Section (\ref{fd}) (cf. (\ref{eh}), (\ref{rnt}) and (\ref{rnti})), we introduce
the random variable $\widetilde{W}_{h,n}$ that is identically distributed with $\widetilde{\Delta}_h (t,n) = \widetilde{m}_{t+h} - \widehat{\widetilde{m}}_{n,t,h}$, and the set
\begin{equation}
   {\cal E}^{I}_h (n,N) =  \{\widetilde{\Delta}_h (t,n) ,  t = n, \ldots, N\} \nonumber 
\end{equation}
consisting of $h$-step prediction errors for $h = 1, \ldots, T$. By the similar technique used in (\ref{rnt}), we approximate  ${\cal I}_{n,N, T}$ by
\begin{equation}
\widehat{{\cal I}}_{n, N, T} =   \sum_{j=N+1}^{N+T} \left( \sum_{h=0}^{n+T-j}  \theta_{j+h-1, h} \right)   \left( \widehat{\widetilde{m}}_{n,N,j-N}+ \widehat{\widetilde{W}}_{n, j-N} \right ) \widetilde{\delta}^*_{j-N}  \nonumber 
\end{equation}
where $\widehat{\widetilde{W}}_h (n,h)$ is the random variable having the same distribution as the empirical distribution constructed by the elements of  ${\cal E}^I_h $; and  $\{\widetilde{\delta}^*_{h}, h = 1,\ldots, T\}$ is an independent Rademacher sequence. Let
$\{\widehat{\widetilde{W}}^{(i)}_{n, h}, h = 1, \ldots, T\}$  and  $ \{ \widetilde{\delta}^{*(i)}_{h}, h=1, \dots, T\}$ be an independent copy of $\{\widehat{\widetilde{W}}_{n, h}, h = 1, \ldots, T\}$  and  $ \{\widehat{\widetilde{\delta}}_{h}, h=1, \dots, T\}$, respectively.
We  propose using the empirical distribution formed by the independent samples
\begin{equation}
\widehat{I}^{(i)}_{N, T} =   \sum_{j=N+1}^{N+T} \left( \sum_{h=0}^{n+T-j}  \theta_{j+h-1, h} \right)   \left( \widehat{\widetilde{m}}_{n,N,j-N}+ \widehat{\widetilde{W}}^{(i)}_{n,j-N} \right) \widetilde{\delta}^{*(i)}_{j-N}, \  i= 1, \ldots, B,  \nonumber 
\end{equation}
as the estimate  $\widehat{F}^{I}_{N,T}(x )$  of the conditional distribution   $F^{I}_{N,T}(x )$  of  ${\cal I}_{n, N,T}$ . Combining (\ref{snt}) and the preceding derivation of  $\widehat{F}^{I}_{N,T}(x )$ yields the desired forecast   $T\mu+ {\cal T}_{n, N,T} +  \widehat{F}^{I}_{N,T}(x )$  of  the conditional distribution of of $S_{N,T}$.
Replace $\mu$ by the sample mean $\widehat{\mu}$ if $\mu$ is unknown.


\section{Non-symmetric shocks}

While the symmetry assumption on the shocks $\{u_t\}$ in model (\ref{model}) is quite common in studies concerning the conditional heteroscedastic model (Christian Francq and Jean-Michel Zakoian 2010), many works also point out that using the non-symmetric shocks such as the skewed normal or skewed-t can bring some performance improvements (see, e.g., Dongming Zhu and John W. Galbriath, 2010,  2011, and references there in). In this section we discuss how to extend our prediction procedure to the case where $\{u_t\}$ is not symmetric. We first focus on the GV model with $r_t = \mu + h_t u_t$ where the iid zero-mean-unit-variance shock sequence $\{u_t\}$ need not be symmetric.  Define $r^*_t = r_t - \mu = m_t \delta_t$ with $m_t = |r^*_t|$. For a small $\lambda >0$, we discretize $r^*_t $ as  $r^*_{t, \lambda} = m_{t, \lambda} \delta_t$ where
\begin{equation}
m_{t, \lambda} = j  \lambda  \quad \mbox{  if  }  m_t \in   I_{j, \lambda} = [j \lambda,  (j+1) \lambda ), \quad  j = 0, 1, \ldots. \nonumber
\end{equation}
Let  $X_{\lambda}$ be a random variable having the same distribution as $r^*_{t, \lambda}$. Conditional on $\{ m_{t,  \lambda}, t =1, \ldots, T  \}$, define

\[
\delta'_{t, \lambda } (m_{t, \lambda} = j\lambda) = \left\{ \begin{array} {ll}
1 & \quad \mbox{with  probability} \quad P ( X_ {\lambda)} \in I_{j, \lambda} ) / P (|X_ {\lambda}| \in I_{j, \lambda}  )  \\

-1 &  \quad  \mbox{with  probability} \quad P ( X_{\lambda} \in - I_{j, \lambda} ) / P (|X_{\lambda}| \in I_{j, \lambda} )\\
\end{array}\right.
\]
Note that if $u_t$ is symmetric, then the conditional probability that $\delta'_{t, \lambda}(m_t) = 1$  or $-1$ is always $1/2$. Define
\begin{equation}
r*'_{t, \lambda}  = m_{t, \lambda} \delta'_{t, \lambda } (m_{t, \lambda} ) .
\end{equation}
Then it is not difficult to see that  $r*'_{t, \lambda} $ and $r*'_{t, \lambda} $ have the sme distribution. Since as $\lambda \rightarrow 0$,  $r*_{t, \lambda} $ converges to
$r*_{t}$ for each $t$, so dose ther distribution of $r*'_{t, \lambda} $ to that of  $r_t$.





\vskip1cm

\noindent {\bf  References}\\

\noindent Bollerslev, T. (1986). Generalized autoregressive conditional heteroscedasticity. \textit{Journal of
    Econometrics} \textbf{31}, 307-327. \\

\noindent Boudoukh, J., Richardson, M., and Whitelaw, R. F. (2008). The myth of long-horizon predictability. \textit{
Review of Financial Studies } \textbf{ 21}, 1577–1605.\\

\noindent Chung, K. L. (2001). \textit{A Course in Probability}. San Diego: Academic Press. \\

\noindent Christoffersen, P.F. (1998). Evaluating interval forecasts. \textit{International Economic Review}
\textbf{39}, 841-862.\\

\noindent Engle, R. and Manganelli, S. (2004). CAViaR: Conditional autoregressive value at risk by regression
quantiles. \textit{Journal Business and Economics and Statistics} \textbf{22}, 367-381. \\

\noindent Fama, E. F and French, K. R. (2018). Long-horizon returns. \textit{ Review of Asset Pricing Studies },
forthcoming. Available at  https://academic.oup.com/raps/advance-article/doi/10.1093/rapstu/ray001/4810768 \\

\noindent Ho, H.-C. (2017). A Non-parametric Estimate of Conditional Tail Expectation for Non-stationary Processes.
Working paper for the grant supported by the Ministry of Science and technology (106-2118-M-001-008).\\

\noindent  Ho, H.-C., Chen, H. and Tsai,  H. (2016). Value at risk for integrated returns
and its applications to equity portfolios. \textit{Statistica Sinica} {\bf 26}, 1631-1648.\\

\noindent Lee, F. K. (2013). Demographics and the long-horizon returns of dividend-yield strategies. \textit{
Quarterly Review of Economics and Finance } \textbf{53}, 202 - 218.\\

\noindent Neuberger, A. (2012). Realized skewness. \textit{ Review of Financial Studies } \textbf{25}, 3423 - 3455. \\

\noindent Taylor, S.J. (1994). Modeling stochastic volatility: A review and comparative study. \textit{Mathematical Finance} \textbf{4}, 183-204. \\


\noindent Artzner, P., Delbaen, F. and Eber, J-M. (1999). Coherent measures of risk. \textit{Mathematical
Fiance} \textbf{9}, 203-228.\\

\noindent Basel Committee on Banking Supervision (2012). \textit{Consultative Document: Fundamental Review of the
Trading Book, Basel, Switzerland.} Available at:

\noindent http://www.bis.org/publ/bcbs219.pdf\\

\noindent Breidt, F.J., Crato, N. and De Lima, P. (1998). The detection and estimation of long memory in stochastic
volatility. \textit{Journal of Econometrics} \textbf {73}, 325-348.\\

\noindent Hamilton, J. D. (1994). \textit{Time Series analysis}. University Press, New Jersey.\\

\noindent Hardy, M.R. (2001). A regime-switching model of long-term stock returns. \textit{North American
Actuarial Journal} \textbf{5}, 41-53.\\

\noindent Ho, H. C. and Hsing, T. (1997). Limit theorems for functionals of moving averages. \textit{The Annals of
Probability}
\textbf{25}, 1636-1669.\\

\noindent  Ho, H.-C., Yang, S. S. and Liu, F.I. (2010). Evaluating quantile reserve for equity-linked
insurance in a stochastic volatility model: long vs. short memory. \textit{ASTIN Bulletin} \textbf {40}, 669-698.\\

\noindent Hosking, J.R.M. (1981). Fractional differencing. \textit{Biometrika} \textbf{68}, 165-176.\\

\noindent Lobato, I.N. and Savin, N.E. (1998). Real and spurious long-memory properties of stock market data.
\textit {Journal of Business and Economic Statistics} \textbf{16}, 261-268.\\

\noindent  Acerbi, C., Tasche, D. (2002). On the coherence of expected shortfall. \textit{Journal of Banking
and Finance} \textbf{26}, 1487-1503.\\

\noindent Ahmadi-Javid, A. (2011). An information-theoretic approach to constructing coherent risk measures.
St. Petersburg, Russia: Proceedings of IEEE International Symposium on Information Theory. 2125-2127.\\

\noindent Ahmadi-Javid, A. (2012). Entropic value-at-risk: A new coherent risk measure. \textit{Journal of
Optimization Theory and Applications} \textbf{155}, 1105-1123.\\


\noindent Artzner, P., Delbaen, F., Eber, J-M. (1999). Coherent measures of risk. \textit{Mathematical
Fiance} \textbf{9}, 203-228.\\

\noindent Asai, M., McAleer, M., Yu, J. (2006). Multivariate Stochastic Volatility: A Review.
\textit{Econometric Reviews} \textbf{25}, 145-175.\\

\noindent Asimit, A.V., Furmanb, E., Tang Q., Vernic, R. (2011).  Asymptotics for risk capital allocations
based on conditional tail expectation. \textit{Insurance: Mathematics, and Economics} \textbf{49}, 310-324.\\

\noindent Basel Committee on Banking Supervision (2012). \textit{Consultative Document: Fundamental Review of
the Trading Book, Basel, Switzerland.} Available at:

\noindent http://www.bis.org/publ/bcbs219.pdf\\


\noindent Bollerslev, T. (1986). Generalized autoregressive conditional heteroscedasticity. \textit{Journal of
Econometrics} \textbf{31}, 307-327.\\

\noindent Breidt, F.J., Crato, N., and De Lima, P., 1998. The detection and estimation of long memory in stochastic
volatility. Journal of Econometrics 73, 325-348.


\noindent Detlefsen, K., Scandolo, G. (2005).
Conditional and Dynamic Convex Risk Measures. \textit{SFB 649 Discussion Paper 2005-006}. Available at:

\noindent http://edoc.hu-berlin.de/series/sfb-649-papers/2005-6/PDF/6.pdf\\

\noindent Du, Z., Escanciano, J. C. (2015). Backtesting Expected Shortfall: Accounting for Tail Risk.
\textit{CAEPR Working Paper 2015-001.} Available at:

\noindent http://goo.gl/xTYKgM\\

\noindent Drost, F. C., and Nijman, T. (1993), Econometrica,  61  (4), 909-927.

\noindent Engle, R.F. (1982). Autoregressive conditional heteroscedasticity with estimates of the variance of
United Kingdom inflations. \textit{Econometrica} \textbf{50}, 987-1007.\\

\noindent Geman, H., Ohana, S. (2008). Time-consistency in managing a commodity portfolio: A dynamic risk
measure approach.  \textit{Journal of Banking \& Finance} \textbf{32} 1991–2005.\\

\noindent Hamilton, J. D. (1994). \textit{Time Series analysis}. University Press, New Jersey.\\

\noindent Hardy, M.R. (2001). A regime-switching model of long-term stock returns. \textit{North American
Actuarial Journal} \textbf{5}, 41-53.\\

\noindent Hardy, M. (2003). \textit{Investment Guarantees: Modeling and Risk Management for Equity-Linked Life
Insurance}. John Wiley \& Sons, Inc.\\

\noindent Hardy, M, Freeland, R. K., Till, M.C. (2006). Validation of long-term equity return models for
equity-linked guarantees. \textit{ North American Actuarial Journal} \textbf{10}, 28-47.\\

\noindent Harvey, A., Ruiz, E., Shephard, N. (1994). Multivariate stochastic variance models.
\textit{Review of Economic Studies} \textbf{61}, 247-264.\\

\noindent Ho, H.-C. (2017). A Non-parametric Estimate of Conditional Tail Expectation for Non-stationary Processes. Working paper for
the grant supported by the Ministry of Science and technology (106-2118-M-001-008).\\

\noindent  Ho, H.-C., Chen, H., Tsai,  H. (2016). Value at risk for integrated returns
and its applications to equity portfolios. {\it Statistica Sinica} {\bf 26},1631-1648.\\

\noindent Ho, H. C., and Hsing, T., 1997. Limit theorems for functionals of moving averages. The Annals of Probability
25, 1636-1669.\\

\noindent  Ho, H.-C., Yang, S. S., Liu, F.I. (2010). Evaluating quantile reserve for equity-linked
insurance in a stochastic volatility model: long vs. short memory. {\it ASTIN Bulletin} {\bf 40}, 669-698.\\

\noindent Hosking, J.R.M., 1981. Fractional differencing. Biometrika 68, 165-176.\\

\noindent Jacquier, E., Polson, N. G., Rossi, P. E. (1994). Bayesian analysis of stochastic volatility
models. \textit{Jounal of Business \& Economic Statistics} \textbf{12}, 371-389. \\

\noindent Kastner, G. (2016). Dealing with Stochastic Volatility in Time Series Using the R Package stochvol. \textit{Journal of Statistical Software}, 69(5), 1-30.\\

\noindent Kastner, G. Fr\"{u}hwirth-Schnatter, S. (2014). Ancillarity-sufficiency interweaving strategy (ASIS)
for boosting MCMC estimation of stochastic volatility models. \textit{Computational Statistics \& Data Analysis} \textbf{76}, 408-423.\\

\noindent Lai, T. L., Xing, H. (2008). \textit{Statistical models and methods for financial markets}. Springer, New York.\\

\noindent Lobato, I.N., Savin, N.E. (1998). Real and spurious long-memory properties of stock market data.
 {\textit Journal of Business and Economic Statistics} {\textbf 16}, 261-268.\\

\noindent McNeil, A.J., Frey, R., Embrechts, P. (2005). \textit{Quantitative Risk Management.} Princeton
University Press, Princeton, NJ.\\

\noindent Politis, D.N., Romano, J.P., Wolf, M. (1999). \textit{%
Subsampling}. Springer, New York.\\

\noindent Riedel, F. (2004). Dynamic coherent risk measures. \textit{Stochastic Processes and their
Applications}. \textbf{112}, 185-200.\\

\noindent Taylor, S. (1986). \textit{Modelling Financial Time Series}. John Wiley \& Sons, New York.\\

\noindent  Wang, J.-N, Yeh, J.-H, Cheng, N.Y.-P., (2014)  J. Banking and Finance 35(5), 1158-1169.

\noindent  Wang, J.-N., Du, J.  and Hsu, Y.-T.  (2018). J. of Empirical Finance V. 47, 120-138.

\noindent Yu, J., Meyer, R. (2006). \textit{Multivariate Stochastic Volatility Models: Bayesian Estimation and Model Comparison}.
\textit{Econometric Reviews} \textbf{25}, 361-384.\\

\end{document}